\documentclass[twocolumn]{aastex63}

\usepackage{import}
\usepackage{color}
\usepackage{listings}
\usepackage{amsmath}
\usepackage{graphicx}
\usepackage[normalem]{ulem}

\newcommand{\crossout}[1]{\textbf{\sout{#1}}}

\begin{document}

\shortauthors{Ferrer Chávez, Wang \& Blunt}
\shorttitle{Biases in orbit-fitting of directly-imaged exoplanets}

\title{Biases in orbit fitting of directly-imaged exoplanets with small orbital coverage}

\author{Rodrigo Ferrer Chávez}
\affiliation{School of Engineering, Autonomous University of Yucatán, Mérida, Yucatán, México}

\author{Jason J. Wang}
\affiliation{Department of Astronomy, California Institute of Technology, Pasadena, CA, USA}
\affiliation{51 Pegasi b Fellow}
\affiliation{Denotes Equal Coauthorship}

\author{Sarah Blunt}
\affiliation{Department of Astronomy, California Institute of Technology, Pasadena, CA, USA}
\affiliation{NSF Graduate Research Fellow}
\affiliation{Denotes Equal Coauthorship}

\begin{abstract}
    The eccentricity of a planet's orbit and the inclination of its orbital plane encode important information about its formation and history. However, exoplanets detected via direct-imaging are often only observed over a very small fraction of their period, making it challenging to perform reliable physical inferences given wide, unconstrained posteriors. The aim of this project is to investigate biases (deviation of the median and mode of the posterior from the true values of orbital parameters, and the width and coverage of their credible intervals) in the estimation of orbital parameters of directly-imaged exoplanets, particularly their eccentricities, and to define general guidelines to perform better estimations of uncertainty. For this, we constructed various orbits and generated mock data for each spanning $\sim 0.5 \%$ of the orbital period. We used the Orbits For The Impatient (OFTI) algorithm to compute orbit posteriors, and compared those to the true values of the orbital parameters. We found that the inclination of the orbital plane is the parameter that most affects our estimations of eccentricity, with orbits that appear near edge-on producing eccentricity distributions skewed away from the true values, and often bi-modal. We also identified a degeneracy between eccentricity and inclination that makes it difficult to distinguish posteriors of face-on, eccentric orbits and edge-on, circular orbits. For the exoplanet-imaging community, we propose practical recommendations, guidelines and warnings relevant to orbit-fitting.
\end{abstract}


\section{Introduction}
\label{sec:intro}

The orbit of a planet can reveal much about its history and properties. For example, distinct population-level eccentricity distributions have been used to suggest different formation mechanism between giant planets and brown dwarfs \citep{Bowler_2020}. Planets detected via the radial velocity (RV) technique also exhibit different population-level eccentricities between long-period and short-period planets \citep{Kipping_2013}. Characteristics of the orbit's eccentricity can also be used to propose the presence of unseen planets, like in the case of Fomalhaut b \citep{Kalas_2013,Faramaz_2015}. Additionally, the inclination of the orbital plane relative to the stellar spin axis can be used as evidence of disturbances during the planet's formation \citep{Maire_2019, Kraus_2020, Bryan_2020}. 

Direct-imaging has allowed us to detect planets at wider separations than indirect detection methods \citep{Bowler_2016}. However, these wide separations often imply very long orbital periods of hundreds to thousands of years. As a result, even after years of observations, often times we only see a given planet moving along a very small arc of its orbit, making orbit-fitting challenging and potentially susceptible to systematic or statistical biases. 

In the case of systematics, one possible cause for biased estimates is using data from different sources. Different observing strategies or different calibrations can significantly alter the results of the orbit-fit \citep{Millar_Blanchaer_2015}. Furthermore, as highlighted by \citet{Konopacky_2016}, systematic errors from different data-reduction pipelines can produce noticeable biases in the estimation of orbital elements, which can persist even after attempts to mitigate them.

On the other hand, statistical biases can arise from the use of Bayesian inference, which is a common tool in determining orbits. Bayesian orbit-fitting is the process of taking measurements of a companion's position relative to its primary and converting them to posterior probability distributions over its orbital parameters. Bayesian methods have been widely used in orbit-determination because they produce full posteriors and allow us to properly assess uncertainties, instead of just producing families of best-fit solutions. Bayesian orbit-fitting is especially useful when relatively little data is available, because the prior probability distributions of Bayes' theorem enable constraints on orbital parameters based on geometric or physical properties of orbits. However, with very limited orbital coverage, biases in the posteriors due to the choice of priors for each orbital element become a concern, since the prior can easily dominate over the likelihood \citep{Lucy_2014}. The current standard practice in the field is to choose uniform or ``uninformative" priors, but little research has been devoted to the consequences of these assumptions. In this regard, \citet{O_Neil_2019} proposed the alternative of observable-based priors to mitigate biases and prevent the under-estimation of credible intervals. Their priors performed well with the orbits of low-phase coverage objects they tested. The effectiveness of their proposed priors for arbitrary orbits is yet to be evaluated through a systematic study of parameter space. 

Another possibility to mitigate the problem of biases might be to explore prior-independent approaches to constrain aspects of the orbit or to constrain different elements of the orbit. One such example was proposed in \citet{Pearce_2015} for very short orbital arcs, especially when orbital curvature is not observed. They provided ways to bound certain orbital elements in a prior-independent manner (though implicitly assuming uniform priors in the line of sight position and velocity) and showed an alternative way to visualize the set of possible orbits in terms of the line of sight position and velocity. Nevertheless, as was highlighted in the study, their approach only makes use of one single position and velocity, so it does not incorporate the information and constraints provided by orbital curvature, substantial changes in separation while the companion is on linear motion, or additional radial velocity data, so using their method alongside common-practice Bayesian inference techniques is preferable. Still, looking for different ways of parametrizing orbital parameters is an approach worth exploring, since they might be less prone to the biases found when fitting the classic Keplerian elements.

Biases in orbit-fitting can also be caused by degeneracies of different orbital elements with each other, or simply by the complex shape of the multidimensional posteriors. One example of this is 51 Eridani b \citep{DeRosa_2019}, where it was discussed that even though the marginalized posteriors of 51 Eri b suggest that circular orbits can be excluded at high significance, further analyses show that they can still be a good fit for the data. Therefore, further research is needed to identify and understand these cases.

The aim of this paper is to investigate biases in orbit-fitting of directly imaged exoplanets when we have a low orbital coverage. We explore biases in eccentricity posteriors in various regions of parameter space in order to find patterns and possible work-arounds. We discuss our method in Section \ref{sec:methodology}, our results for eccentricity biases in Section \ref{sec:results}, a brief analysis on inclination in Section \ref{sec:inclination}, an inverse analysis concerning measured eccentricities and inclinations in Section \ref{sec:inverse} and case studies as examples of how our results might be applicable in Section \ref{sec:casestudy}. We end with conclusions and practical recommendations in Section \ref{sec:recommendations}.

\section{Methodology}
\label{sec:methodology}

We generated simulated data for a grid of orbits whose orbital parameters we know and used those data to perform orbit fits.

A Keplerian orbit can be defined using 8 elements: semi-major axis ($a$), eccentricity ($e$), inclination ($i$), argument of periastron of the planet ($\omega$), longitude of the ascending node ($\Omega$), epoch of periastron passage expressed as a fraction of the orbital period past a reference epoch ($\tau$), parallax ($\pi$) and total system mass ($m_{tot}$), all as defined in \citet{Blunt_2020}. To see which orbits make it more difficult to estimate eccentricity, given an initial orbit, we performed a one-dimensional (i.e. varying one orbital element at a time) analysis of the orbital parameters: we varied individually the four parameters that affect the shapes of the posteriors ($e$, $i$, $\omega$ and $\tau$) over their ranges in equal steps while keeping the rest fixed ($a=36.44$ AU, $\Omega=248.66 ^\circ$, $\pi = 56.89$ mas, $m_{tot} = 1.28 M_{\odot}$). The values used for the four chosen parameters are indicated in the first four rows of Table \ref{tab:testedparams}. 

\begin{deluxetable*}{cccc}
\tablecaption{Tested orbital parameters$^a$ \label{tab:testedparams}}
\tablehead{\colhead{Test} & \colhead{Orbital element} & \colhead{Tested values} & \colhead{Number of orbits generated}}
\startdata
1-D & $e$& $0$ to $1$ in steps of $0.1$ & $11$\\
\tableline
1-D & $i$& $0^\circ$ to $180^\circ$ in steps of $18^\circ$ & $11$ \\ 
\tableline
1-D  & $\omega$& $0^\circ$ to $360^\circ$ in steps of $36^\circ$ & $11$ \\
\tableline
1-D & $\tau$& $0$ to $1$ in steps of $0.1$ & $11$ \\
\tableline
2-D &$e$ & $0$ to $0.9$ in steps of $0.1$ & $110$\\
& $i$& $0$ to $180^\circ$ in steps of $18^\circ$&\\
\tableline
3-D &$e$& $0$ to $0.9$ in steps of $0.1$, and $e=0.99$ & $396$\\
& $i$ & $0^\circ$ to $90^\circ$ in steps of $18^\circ$ &\\ 
& $\tau$& $0$ to $0.8$ in steps of $0.2$, and $\tau = 0.3$ &\\
\tableline
Second 1-D & $e$& $0$ to $1$ in steps of $0.1$& $11$\\
\tableline
Second 1-D & $i$ & $0^\circ$ to $180^\circ$ in steps of $18^\circ$ & $11$
\enddata
\tablenotetext{a}{Orbital parameters used for the tests described in Section \ref{sec:methodology}. In the one-dimensional tests the parameters were varied individually, while in the two- and three-dimensional tests they were varied jointly in pairs and triplets, respectively. The second one-dimensional test (last two rows of the table) uses four data points evenly spaced in time instead of the three used in the rest of the tests. The values of the fixed parameters are specified in the text.}
\end{deluxetable*}

When other individual parameter was being varied, the parameters were fixed to $e = 0.33$, $i = 153.62^{\circ}$, $\omega = 251.99^{\circ}$ and $\tau = 0.46$. We did not perform tests with the rest of the orbital parameters since they do not alter the shape of the orbit; for example, varying semi-major axis only changes the size of the orbit, and varying the longitude of the ascending node only rotates it around our line of sight. 

Having specified each set of orbital parameters, we generated mock data for each of them. Given the orbital elements and the desired observation epochs, we solved Kepler's equation to get the positions of the planet relative to the primary's location in right ascension ($\Delta$RA) and declination ($\Delta$Dec) for those epochs. Then we added noise to each data point in both $\Delta$RA and $\Delta$Dec, drawing values from a normal Gaussian distribution centered at $0$ $\mathrm{mas}$ with $\sigma = 1$ $\mathrm{mas}$, which is a level of precision comparable to real-life astrometry (e.g. \citealt{Wang_2016}). 

\begin{figure*}
\includegraphics[width=0.44\textwidth]{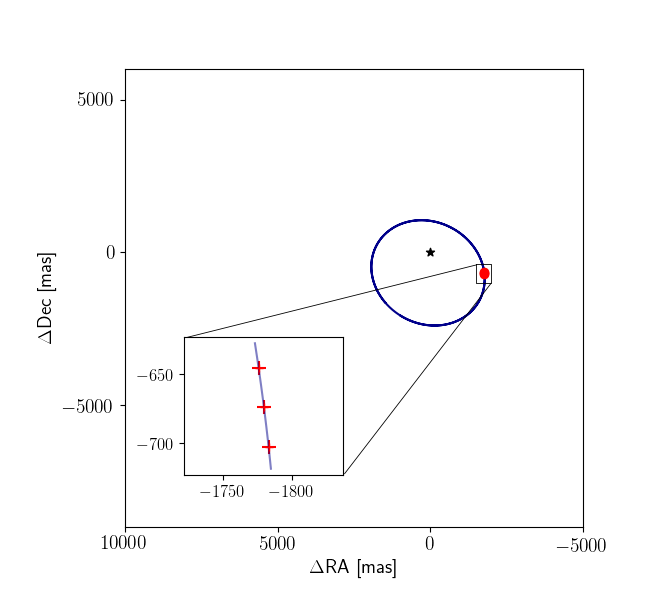}
\includegraphics[width=0.50\textwidth]{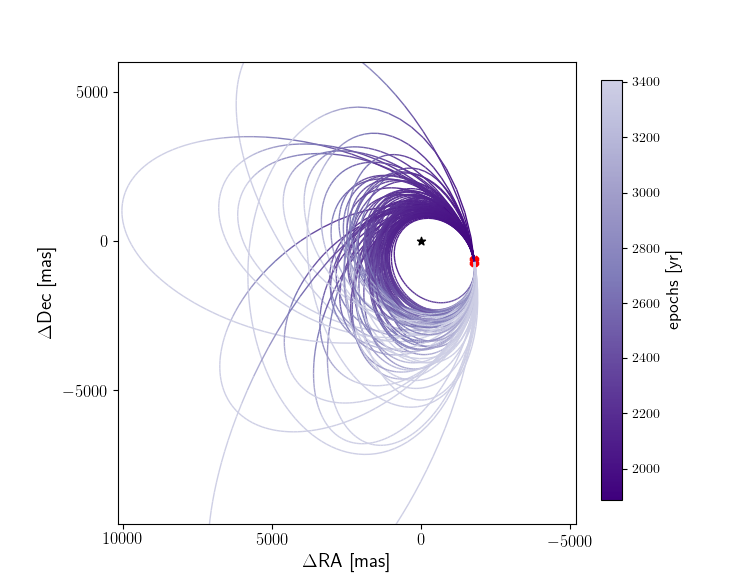}
\caption{Illustration of the process for generating mock data. Left: Initial sky-projected orbit from which the three ideal data points were drawn, spanning $\sim 0.48 \%$ of the total orbital period. Gaussian noise was added to mimic measurement uncertainties. Right: Orbit posteriors calculated with \texttt{orbitize!}. These were generated using the OFTI algorithm on the previously mentioned mock data, obtaining $10,000$ total orbit fits for each particular data set. In both panels, the data are represented by red crosses and the position of the primary is indicated with a black star at $(0,0)$.}
\label{fig:fakedata}
\end{figure*}

Since the problem of biases is the strongest when we have very little orbital coverage, for all tests we generated three data points equally spaced in time spanning $0.48 \%$ of the total orbital period. An orbital coverage of less than $1\%$ of the period was chosen because in this regime, where not even orbital acceleration can be measured in most cases, we can get very fast orbit-fits with the Orbits For The Impatient (OFTI) Bayesian rejection sampling algorithm \citep{Blunt_2017}, and also because many planets in real life have been observed with similarly small or smaller orbital coverage (e.g. HD 23514 B, 2M1559+4403 B and 1RXS2351+3127 B, according to the best-fit orbital periods reported by \citealt{Bowler_2020}). We used the OFTI algorithm to fit the orbits of the simulated data with the \texttt{orbitize!} package using its default (uninformative) priors for the orbital elements, all specified in Table 1 of \citealt{Blunt_2020}.We obtained 10,000 accepted orbits in each case. The process of generating orbit fits from simulated data is visualized in Figure \ref{fig:fakedata}.

After looking at the results from these one-dimensional tests, a two-dimensional test was run varying eccentricity and inclination simultaneously while fixing the rest of the orbital parameters. The values used for each element are shown in the fifth row of Table \ref{tab:testedparams}. 

In a similar manner, we also looked at the three-dimensional configurations of eccentricity, inclination and $\tau$. The values used are described in the sixth row of Table \ref{tab:testedparams}. In this case, to save computation time, inclination was only varied in the range $0^\circ$ to $90^\circ$ (in the \texttt{orbitize!} coordinate system, $i = 0^{\circ}$ is a face-on orbit and $i=90^{\circ}$ is an edge-on orbit) because orbits inclinations symmetric to $90^\circ$ do not differ in their shapes, but only in going in the clockwise or counterclockwise directions. Also, we originally tested $5$ evenly spaced values of $\tau$, but we decided to test additionally $\tau = 0.3$ since in this case that value corresponds to a near-periastron orbital phase, which was previously absent. 

Finally, more one-dimensional tests were done with four data points evenly-spaced in time instead of three (now the samples spanning $0.72\%$ of the orbital period). The values used are shown in the last two rows of Table \ref{tab:testedparams}. 

The results from this short test were compared with the previous one-dimensional posteriors with three data points, to assess the sensitivity of the results on the number of simulated data points and changes in the orbital coverage.

\section{Results of eccentricity bias}
\label{sec:results}

In this Section we summarize our findings regarding the bias in the eccentricity posterior from the various analyses that we carried out. We remind the reader that all the tests in this paper were performed with very small orbital coverage ($\sim 0.5\%$), where generally orbital acceleration is not measured, so the results and conclusions obtained here might not hold when more data or better data is available.

\subsection{One-dimensional analysis}
\label{sec:analysis1d}

\begin{figure*}
    \centerline{\includegraphics[width=1.1\linewidth]{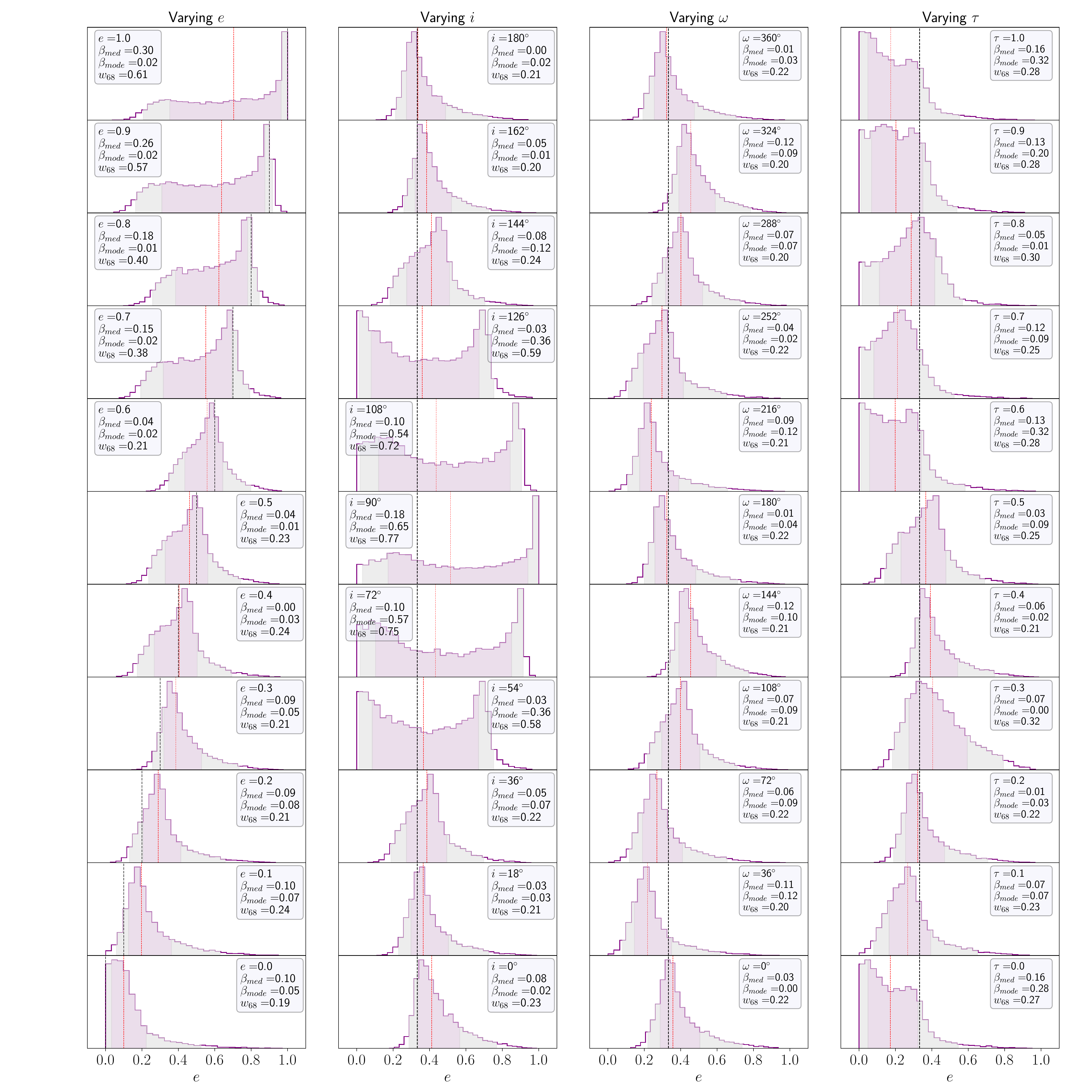}}
    \caption{Eccentricity posteriors from the one-dimensional analysis. The dashed vertical black lines indicate the true value of eccentricity, the dashed vertical red lines indicate the median of the posterior distribution, the purple-shaded region corresponds to the $68\%$ credible interval and the grey-shaded region the $95\%$ credible interval. Each panel has a text box with the used value of the corresponding orbital parameter and biases defined as the absolute difference of the median of the posterior and the true eccentricity ($\beta_{med} = \lvert e_{\mathrm{med}} - e_{\mathrm{truth}} \rvert $), the absolute difference of the mode of the posterior and the true eccentricity ($\beta_{mode} = \lvert e_{\mathrm{mode}} - e_{\mathrm{truth}} \rvert $), and the width of the $68\%$ credible interval $w_{68}$. From left to right: Different eccentricities, inclinations, arguments of periastron and epoch of periastron passage. In this orbit values of $\tau$ from $0.6$ to $1.0$ (or $0.0$, equivalently) correspond to the more near-apastron planet positions. We see that eccentricity and inclination are the orbital elements that affect the most the eccentricity posterior. The biggest effect is seen at near edge-on ($i\sim 90^\circ$) inclinations.}
    \label{fig:stacked_hist}
\end{figure*}

All the eccentricity posteriors from this analysis are shown in Figure \ref{fig:stacked_hist}. We looked at multiple ways to assess the bias present in each orbit-fit. The metrics considered were the width of the $68\%$ credible interval and the absolute difference of the median and mode (binning the data into 30 bins) of the posterior and the true value. The purpose of these is both to quantify bias and to evaluate where some estimators are generally better than others. In each panel of Figure \ref{fig:stacked_hist} we include the three metrics that we considered the most useful, being the following:

\begin{equation}
    \beta_{med} = \lvert e_\mathrm{med} - e_\mathrm{truth} \rvert
\end{equation}

\begin{equation}
    \beta_{mode}= \lvert e_\mathrm{mode} - e_\mathrm{truth} \rvert
\end{equation}

\begin{equation}
    w_{68} = \text{Width of the 68\% credible interval}
\end{equation}

where the credible interval in centered at the median of the eccentricity posterior. The $68\%$ credible interval is defined as the interval comprised between the $16^{th}$ and the $84^{th}$ percentiles of the distribution.

In the first metric we are measuring the absolute difference of the median of the posterior distribution and the true eccentricity, in the second the absolute difference of the mode of the posterior distribution and the true eccentricity, and in the third we are measuring the uncertainty suggested by the posterior. Note that mode bias might change due to the randomness of the process and the choice of bins, since (as can be seen in Figure \ref{fig:stacked_hist}) for certain orbits the eccentricity posterior is bi-modal, having two local peaks near the extremes of similar height, so small changes could cause the mode to choose one peak over the other. To confirm that our posteriors were not merely the result of shot noise, we re-ran all the orbit fits in this section a second time and compared them with the ones presented here. The results were almost identical\crossout{:} \textbf{.} the median difference (and the associated $68\%$ credible interval of said difference) between the medians of the eccentricity posteriors was $0.028^{+0.018}_{-0.020}$, the median difference between the modes was $0.024^{+0.018}_{-0.022}$ and the median difference in the widths of the $68\%$ credible intervals was $0.008^{+0.016}_{-0.003}$. This gives us more confidence about using the specified bias metrics.

When varying true eccentricity from $0$ to $1$ an interesting effect was observed, namely that for moderate and high eccentricities the median was strongly biased towards lower values, while the mode of the distribution remained constantly aligned with the true eccentricity. In the case of $e= 0.9$, for example, the bias of the median was $\beta_{med} = 0.26$, while the bias of the mode was $\beta_{mode} = 0.02$. This hints that the mode might be a better estimator for eccentricity than the median for high eccentricity orbits, so we will keep comparing these two metrics throughout this study. Additionally, the width of the $68\%$ credible interval remained roughly constant from $e = 0$ to $e = 0.6$, at around $w_{68} = 0.21$, and the true value of eccentricity remained within the $68\%$ credible intervals for most of the trials, but at high eccentricities the width of the credible interval increased significantly, becoming wider as the orbit became more eccentric, peaking at $w_{68} = 0.61$ when $e = 1$. 

In the case of different inclinations, for face-on orbits eccentricity was well-constrained, but that was not the case for near-edge-on orbits: not only are they notably wide (the widest being $w_{68} = 0.77$ for a perfectly edge-on orbit) and bi-modal, but their peaks are extremely distant from the true value of $e$ (the furthest being $\beta_{mode} = 0.65$ with the $i = 90^\circ$ orbit in the second column of Figure \ref{fig:stacked_hist}). It is also interesting to note that the mode of the distribution was very good at estimating eccentricity for face-on inclinations, but very bad at edge-on inclinations. Also, it is worth noting that the bias of the median is very small and that the true eccentricity landed within the $68\%$ credible interval in most cases, but this might be simply due to the fact that we chose a moderate true eccentricity (we set $e = 0.33$ for the three right-most columns of Figure \ref{fig:stacked_hist}), so different true eccentricities at near-edge-on inclinations will be tested in the following sections. This heavy dependency of eccentricity in inclination was the most significant one from the four parameters that we tested, so it will be one of our main focuses of discussion as we go to higher dimensions.

The posterior eccentricity distribution remains mostly unchanged for all values of $\omega$, being only slightly displaced from the true value of $e$ in almost all tests. In some cases the true value of eccentricity lands outside the $68\%$ credible interval, but the width of the interval remains mostly constant at $\sim 0.21$ when varying $\omega$. Furthermore, the median and the mode themselves were very close to each other and to the true eccentricity: the biggest bias of the median was $\beta_{med} = 0.12$ and the biggest bias of the mode was $\beta_{mode} = 0.12$ as well. We took this behavior as evidence that the argument of periastron does not play a significant role in estimating eccentricity.

For different values of $\tau$ the eccentricity posterior was also affected, but not to a great extent. In some cases the distribution was skewed towards lower values (the biggest bias of the median was $\beta_{med} = 0.16$). The width of the $68\%$ credible interval changed from a minimum of $w_{68} = 0.21$ to a maximum of $w_{68} = 0.32$, which is not huge compared to the credible intervals that we found at near-edge-on inclinations. For a meaningful conclusion about the dependence on orbital phase we need to look at the multidimensional tests, which we will do in Section \ref{sec:analysis3d}.

The key takeaways from this section are the following:
\begin{itemize}
    \item The eccentricity posterior is the most sensitive to $e_{truth}$ and $i_{truth}$, while it is the least sensitive to $\omega$,
    \item The eccentricity posterior gets skewed towards lower values at high values of $e_{truth}$,
    \item Near edge-on values of $i_{truth}$ produce bi-modal eccentricity posteriors.
\end{itemize}

\subsection{Two-dimensional analysis}
\label{sec:analysis2d}

\begin{figure*}
    \centerline{\includegraphics[width=1.1\textwidth]{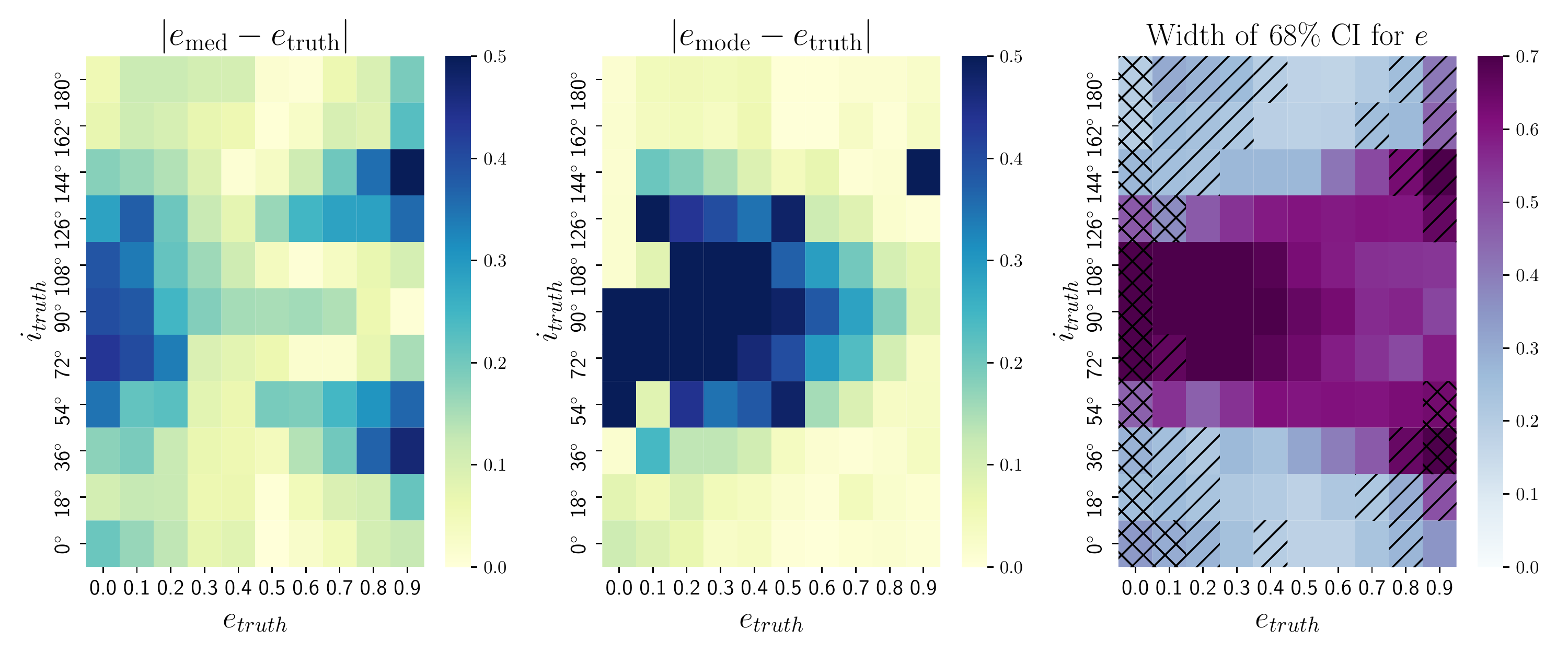}}
    \caption{Biases resulting from varying both inclination and eccentricity simultaneously. In this test, the planet position was about midway in position between periastron and apastron ($\tau = 0.46$). Each cell corresponds to a single orbit whose true eccentricity and true inclination are given by the $x$ and $y$ axis, respectively. Each cell is colored according to the magnitude of the specified bias metric. Cells hatched with a "//" pattern indicate orbits where $e_{truth}$ landed outside the $68\%$ credible interval, and cells hatched with an "X" pattern indicate orbits where $e_{truth}$ landed outside the $95\%$ credible interval.  Left: Bias is measured by the absolute difference of the median and the true  eccentricity. Middle: Bias is measured by the absolute difference of the mode and the true value of eccentricity. We capped the maximum value of the color bar for the median and the mode at $0.5$. Right: Bias is measured by the width of the $68\%$ credible interval centered at the median. We see that across eccentricities, near-edge-on inclinations produce significant biases for all three metrics.}
    \label{fig:medmodebias}
\end{figure*}

Given that the eccentricity posteriors were by far the most sensitive to the underlying values of $i$ and $e$, it makes sense to look at different pairs of them to identify patterns. After varying inclination and eccentricity simultaneously, we computed the biases of the median ($\beta_{med}$) and the mode ($\beta_{mode}$), as well as the width of the $68\%$ credible intervals centered at the median ($w_{68}$) in each orbit-fit. We present the results as heat maps in Figure \ref{fig:medmodebias}, where darker colors correspond to greater values of the specified metric. We see a clear symmetry along the $i = 90^{\circ}$ row, which is what we would expect, since the only difference of two orbits with inclinations symmetric to $90^{\circ}$ is that one goes in the clockwise direction and the other in the counter clockwise direction. 

Once again, to make sure the general patterns in our posteriors were not merely the result of noise, we re-ran the orbit-fits in this two-dimensional test and compared the results with the ones presented here. Like in the one-dimensional tests, the results were almost identical: the median difference (and the associated $68\%$ credible interval of said difference) between the medians of the eccentricity posteriors was $0.024^{+0.035}_{-0.016}$, the median difference between the modes was $0.016^{+0.025}_{-0.015}$ and the median difference in the widths of the $68\%$ credible intervals was $0.022^{+0.026}_{-0.015}$. 

We find that the mode bias is largest for values of inclination around $90^{\circ}$ for most values of eccentricity. This is consistent with the posteriors of the uni-dimensional analysis of Figure \ref{fig:stacked_hist}, where the same pattern was first suggested. In fact, by comparing "varying $i$" column in Figure \ref{fig:stacked_hist} with the columns of constant true eccentricity in Figure \ref{fig:medmodebias} it may be easier to understand some characteristics of the heat map. For instance, the mode bias at edge-on inclinations is less at very high eccentricities: this is simply because these posteriors present a peak near $e \sim 1$ independently of the true value of $e$. This can also help us understand some of the irregularities in the $\beta_{mode}$ grid, that is, apparent big jumps in bias like in the cells $e_{truth} = 0.1$, $i_{truth} = 34^\circ$ and $e_{truth} = 0.9$, $i_{truth} = 144^\circ$. These stem due to bi-modal posteriors, like we discussed in Section \ref{sec:analysis1d}. Still, the previously mentioned symmetry about $i_{truth} = 90^{\circ}$ in the Middle panel of Figure \ref{fig:medmodebias} gives us confidence about using $\beta_{mode}$ as a consistent measure of bias, regardless of the irregularities in the $\beta_{mode}$ grid due to bi-modal posteriors.

A similar argument can be made for the median bias diminishing at moderate values of $e$: the median of those posteriors is located around the center of the eccentricity range independently of the true value of $e$, so that when $e_{truth}$ happens to be near the center it will be close to the median, contrary to the case when $e_{truth}$ is near the edges of the eccentricity range. This is consistent with the $w_{68}$ grid and the distribution of orbits with $e_{truth}$ within the $68\%$ credible intervals (cells without hatch in Figure \ref{fig:medmodebias}), since we see that the $68\%$ credible intervals are very wide for all eccentricities at near-edge-on inclinations compared to near-face-on inclinations, including the lighter regions of the $\beta_{med}$ grid. Thus, smaller biases in the median or orbits with true values within the $68\%$ credible interval in this regime do not necessarily translate into well-constrained eccentricities.

Regarding the discussion of credible intervals it is worth noting, however, that when the posterior is near the edges of the allowed parameter space (i.e. $0$ or $1$ for eccentricity) in practice an upper or lower $68\%$ limit would be reported instead of the credible intervals centered at the median, since the latter will naturally exclude values near the edges. In this paper, for simplicity, we only analyze the credible intervals.

In practice, considering the three grids in Figure \ref{fig:medmodebias}, we can say that, in near-edge-on orbits, eccentricity will be essentially unconstrained, since neither the median nor the mode map one-to-one to the true underlying values of $e$, and the credible intervals are notably wide in all cases. Still, the wide credible intervals do reflect the breadth of the uncertainty. 

One interesting thing to note is that in the regions near $i = 0^{\circ}$ or $i = 180^{\circ}$ (that is, near-face-on orbits) the mode bias is consistently lower than the median bias. Additionally, near-face-on, near-zero eccentricity regions in the $\beta_{med}$ grid exhibit a noticeable bias and the true values are constantly outside the $68\%$ credible intervals, but the same regions in the $w_{68}$ grid do not show very wide credible intervals. For this reason, it is likely that the biases of the median in the face-on, circular orbits regime is due to the Lucy-Sweeney bias \citep{lucybias}, that is, simply by its nature the median will fail to reach the lower limit of the eccentricity range. In this inclination regime, even in moderate values of $e_{truth}$, the mode of the eccentricity distribution tends to be closer to the true value than the median overall. A similar case can be made for high eccentricities in almost all inclinations. This suggests that for near-face-on orbits the peak of the eccentricity posterior is a better estimator than the median independently of the value of $e$. This is to be explored in the following subsection.

One remarkable fact is that the mode bias is very strong at near-edge-on values for moderate and low eccentricities, and they gradually diminish with very high eccentricities. This suggests that near-edge-on inclinations heavily bias the eccentricity posteriors towards higher values, which is to be expected from physical intuition: the shape of a face-on eccentric orbit is very similar to a near-edge-on almost circular orbit. The similarity in these shapes is exacerbated by the short orbital arc that we observe: we have data from such a small portion of the orbit that we don't even see changes in the planet's speed, which could give more information to undo this degeneracy.

Additionally, there is indeed symmetry in Figure \ref{fig:medmodebias} with respect to $i = 90^{\circ}$. This is to be expected, since orbits with inclinations symmetric to $90^{\circ}$ don't differ in their shapes, but only in going in clockwise or anti-clockwise directions. We use this fact in the following section to save computation time, and thus we will present results for inclinations spanning from $i = 0^{\circ}$ to $i = 90^{\circ}$.

The key takeaways from the present section are the following:
\begin{itemize}
    \item Near edge-on values of $i_{truth}$ produce biased eccentricity posteriors independently of the underlying $e_{truth}$ (that is, wider $68\%$ credible intervals and larger biases in the median and mode),
    \item At near face-on values of $i_{truth}$, the mode of the eccentricity posterior is closer to $e_{truth}$ than the median.
\end{itemize}

\subsection{Three-dimensional analysis}
\label{sec:analysis3d}

\begin{figure*}
    \centerline{\includegraphics[width=1.2\linewidth]{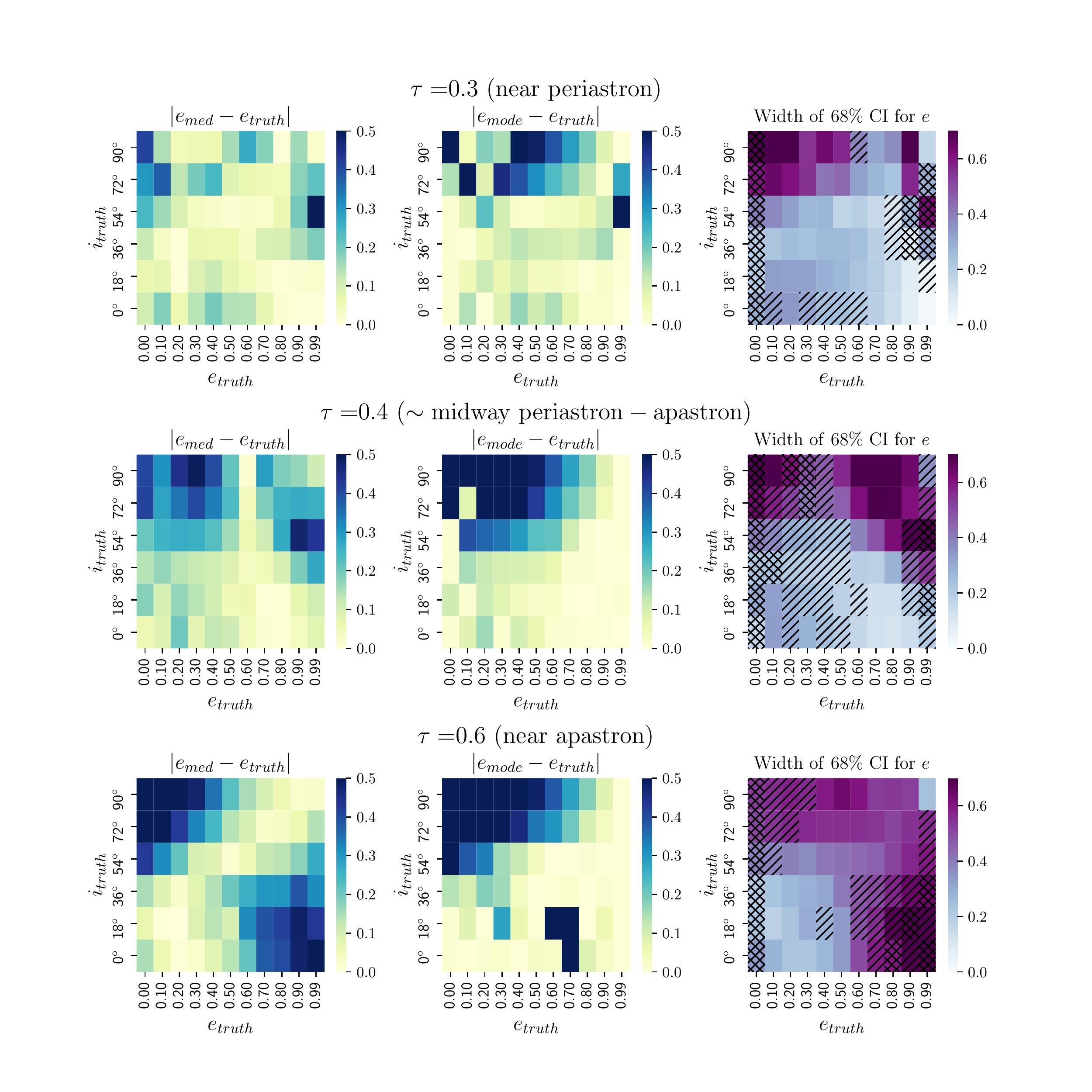}}
    \caption{Same as Figure \ref{fig:medmodebias} but at three different orbital phases. Each row presents the eccentricity biases for the specified orbital phase. Inclination was tested from $0^\circ$ to $90^\circ$. We see that the same basic pattern of Figure \ref{fig:medmodebias} is still present: the eccentricity posteriors have comparably larger biases for all three metrics when the orbit is near-edge-on in most orbital phases. All three biases are also significantly reduced near periastron and enhanced near apastron.}
    \label{fig:fourpanelbias}
\end{figure*}

The same grid of eccentricities and inclinations was simulated for six different planet positions, that is, six different values of $\tau$. Heat maps like the ones in Figure \ref{fig:medmodebias} were generated for each value of $\tau$, three of which are presented in Figure \ref{fig:fourpanelbias} along with the planet position that the used value of $\tau$ represents. We include only those three because symmetric orbital phases produced very similar bias grids.

Like in the two-dimensional analysis, near-edge-on inclinations ($i_{truth} \sim 90^\circ$) produce significant eccentricity biases for all three metrics in most orbital phases. We see that the median bias is very dependent on the position of the planet along the orbit, which is to be expected: since we used a constant time baseline in our analyses, a near-periastron orbital phase translates into faster orbital speed and more fractional orbital coverage compared to non-periastron phases, so in general eccentricity would be better constrained, and thus less biased, when the planet is near periastron. However, near-edge-on inclinations still produce noticeable biases in the median and widen the $68\%$ credible intervals in this regime. 

Similarly, near-apastron orbital phases (which correspond to the slowest orbital speeds and the smallest fractional orbital coverage) have larger biases in the median and wide $68\%$ credible interval for both near-edge-on orbits and very eccentric orbits compared to near-circular, near-face-on orbits. Additionally, the near-circular, face-on ($e_{truth} \sim 0$ and $i_{truth} \sim 0^\circ$) regions of parameter space show very little variation in different orbital phases, which makes sense given that the more circular the orbit, the less difference in orbital speed there is between near-periastron and near-apastron phases.

Near-edge-on inclinations also produce significant biases in the mode for all tested orbital phases. In these inclinations, the bias in the mode seems to be slightly reduced when the planet is near periastron, but all three orbital phases show very similar distributions of the bias of the mode. Importantly, near-periastron orbital phases do not produce huge biases in the mode for very eccentric, near-face-on orbits, contrary to biases in the median. Comparing the biases in the median and the mode, for near-face-on inclinations the mode seems to be at least as good of an estimator for eccentricity as the median, and it becomes significantly better than the median for near-apastron orbital phases. However, the Bottom Middle panel of Figure \ref{fig:fourpanelbias} shows highly biased cells $(e_{truth}, i_{truth}) = (0.70, 0^\circ)$, $(0.70, 18^\circ)$, $(0.60, 18^\circ)$ surrounded by very unbiased cells, and a similar phenomenon takes place in the other mode grids. This indicates that the eccentricity posteriors are bi-modal in these cases, so we need to be careful when assessing whether the mode is preferable to the median in the face-on, high eccentricity, near-apastron regime. In Section \ref{sec:inceccdeg} we will see that the regions of parameter space where we most often see these "outlier cells" across orbital phases (i.e. regions of high eccentricity, face-on inclination and low eccentricity, near-edge-on inclination) are also regions with significant risk of bi-modal posteriors due to the degeneracy between eccentricity and inclination.

Although $\tau$ affects biases and uncertainties, there are certain patterns that hold across orbital phase, the key patterns being:
\begin{itemize}
    \item Both the median and the mode are poor estimators of eccentricity at near-edge-on inclinations,
    \item Near edge-on inclinations produce consistently very wide $68\%$ credible intervals,
    \item The width of the $68\%$ credible interval is usually correlates with the bias present in the median,
    \item Near-face-on, eccentric orbits can produce significant biases in the median and the credible intervals, and
    \item Therefore, there is evidence that for near-face-on inclinations, the mode, in general, is a better estimator than the median of the eccentricity posterior.
\end{itemize}

All the analyses of bias performed in these sections has been relating the posteriors to the underlying orbital parameters, that is, with the knowledge of $e_{truth}$ and $i_{truth}$. In Sections \ref{sec:inverse} and \ref{sec:casestudy} we will see what these results can tell us about the reliability of our posteriors when we do not know the true values of the orbital parameters.

\subsection{One-dimensional analysis with four data points}
\label{fourdata}

When fitting four data points (now spanning $\sim 0.72 \%$ of the orbital period) varying individually eccentricities and inclinations, we found no significant differences in the eccentricity posteriors for the values of $e_{truth}$ and $i_{truth}$ that we tested compared to our previous analysis with three data points. When comparing the posteriors with corresponding values of $e_{truth}$ and $i_{truth}$ with three and with four data points, the median difference (and the associated $68\%$ credible interval of said difference) between the medians of the eccentricity distributions was $0.034^{+0.035}_{-0.018}$, the median difference between the modes was $0.030^{+0.040}_{-0.027}$ and the median difference in the widths of the $68\%$ credible intervals was $0.014^{+0.026}_{-0.008}$.  These variations are very small, actually at the sampling error level (that is, they are the level of variation we would expect from running OFTI multiple times in the same data set). We interpret this as evidence that our results are not sensitive to small changes in orbital coverage.

\section{A look at inclination}
\label{sec:inclination}

\begin{figure*}
    \centerline{\includegraphics[width=1.2\linewidth]{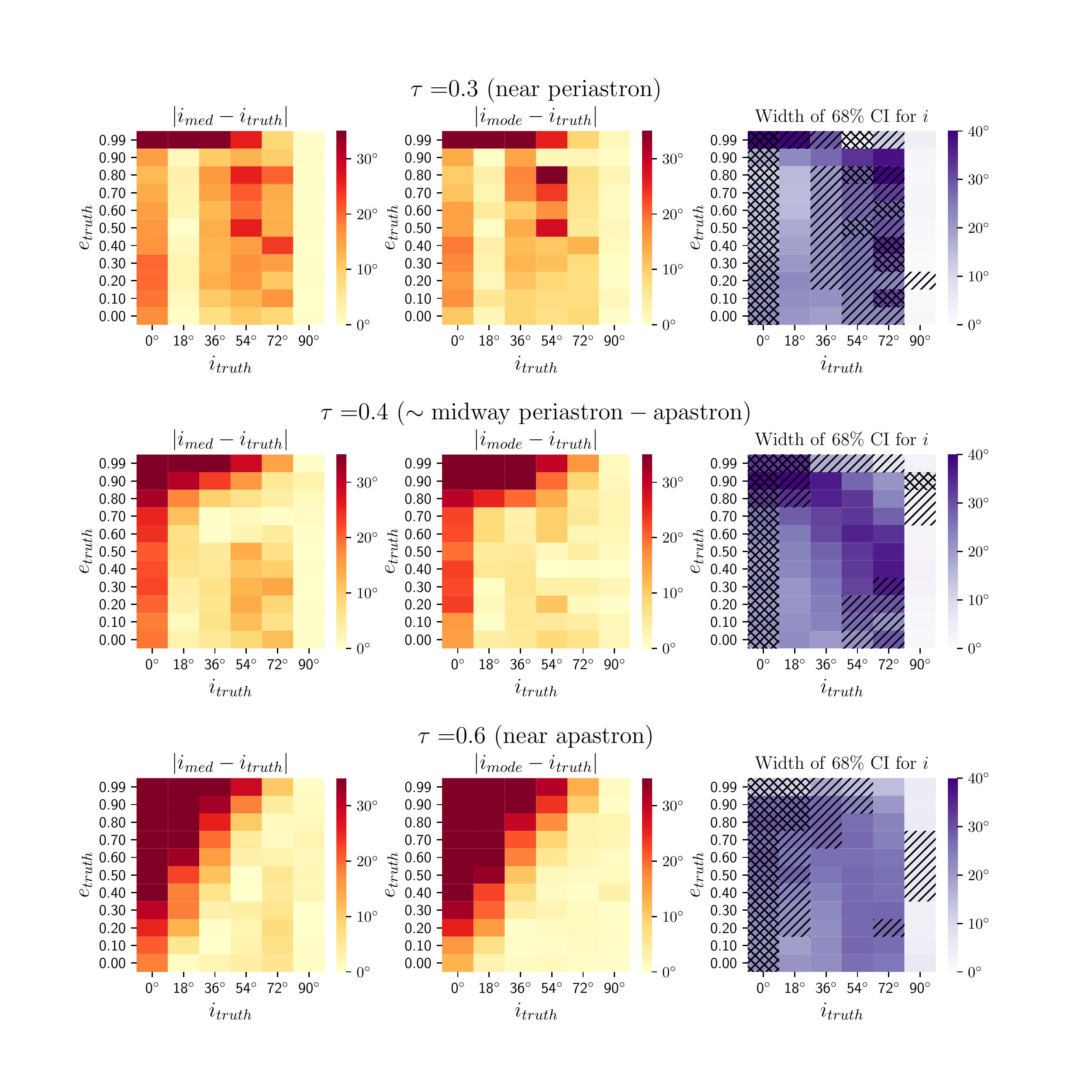}}
    \caption{Inclination biases for different eccentricities, inclinations and planet positions. Each row presents the inclination biases for the specified orbital phase, in degrees. We capped the maximum value of the color bar for the median and the mode grids at $30^\circ$. Cells hatched with a "//" pattern indicate orbits where $i_{truth}$ landed outside the $68\%$ credible interval, and cells hatched with an "X" pattern indicate orbits where $i_{truth}$ landed outside the $95\%$ credible interval. The bias in the median and the bias in the mode follow the same distribution, suggesting that the inclination posteriors most often are not bi-modal.}
    \label{fig:incbias}
\end{figure*}

Given the high dependence of the eccentricity estimates with regards to inclination, we took a look at the inclination posteriors themselves. We present results from the three-dimensional analysis in Figure \ref{fig:incbias}, analogous to the previous ones of eccentricity, also quantifying bias as the absolute difference of the median and the true inclination, the absolute difference of the mode and the true inclination, and the width of the $68\%$ credible interval.

First, we see that across orbital phases the bias in the median and the bias in the mode are consistently following the same distribution, contrary to the case of eccentricity. Additionally, the width of the $68\%$ credible interval did not show any important patterns in the near-apastron phase, and the other two credible intervals were just a little narrower in near-face-on inclinations ($i\sim 0^\circ$). In particular, the distribution of the widths of credible intervals did not follow the distribution of biases in the median. All this together suggests that the inclination posteriors most often were not bi-modal, but they usually pointed to a singular value, regardless of the magnitude of the bias.

Biases are much larger when the planet is near apastron, which is to be expected given the slower orbital speed in this region. In the near-periastron orbital phase we do not see a clear pattern in the distribution of biases in the mode or the median, aside from the most eccentric orbits ($e = 0.99$) being very biased. The rest of parameter space is relatively uniform. It is worth noting, however, that most of the orbital configurations at a near-periastron orbital phase yield posteriors with $i_{truth}$ outside the $68\%$ credible interval (almost all cells are hatched with a "//" pattern), contrary to eccentricity, that has the most orbits with $e_{truth}$ within the $68\%$ credible interval in this orbital phase.

In the second and third rows of Figure \ref{fig:incbias}, we find that increasingly eccentric orbits produce increasingly biased inclination medians and modes, and with them, more orbits with $i_{truth}$ outside the $68\%$ credible interval, though no significant changes in the width of the credible intervals are noticed. The regions with relatively small biases in the median also tend to have $i_{truth}$ within the $68\%$ credible interval. Additionally, the only significant pattern that we find with regards to the $95\%$ credible interval is that, in general, only perfectly face-on orbits have $i_{truth}$ outside the $95\%$ credible interval, aside from a few orbits in the near periastron case with moderate inclinations ($i_{truth} \sim 72^\circ$).  

Looking at the true inclinations and the magnitude of the biases in the median and the mode, we see that the most biased regions in the grids correspond to inclination estimates that are significantly more edge-on than their true inclinations. Thus we find that high-eccentricity orbits bias the median and mode of inclination towards more edge-on values, and that near-edge-on inclinations also bias the median and mode of eccentricity towards higher values, aside from widening the eccentricity credible intervals. Therefore, we find a strong degeneracy between face-on, eccentric orbits and edge-on, circular orbits. We will study the impact this has in our ability to reliably estimate eccentricity and inclination in Sections \ref{sec:inverse} and \ref{sec:casestudy}.

We consistently see that a significant bias of medians and modes is present when $i_{truth} = 0^\circ$ for all eccentricities, even when almost face-on orbits with $i_{truth} = 18^\circ$ have very small biases. This is likely due to the Lucy-Sweeney bias and the choice of $\sin i$ as the prior for inclination. It is worth noting, however, that the shape of the orbit projected onto the sky plane is not as sensitive on small variations on inclination when the orbit is face-on compared to when it is edge-on, since differences in orbit shape on the sky plane are determined by $\cos{i}$ rather than $i$ alone. This means that even though perfectly face-on orbits have the largest biases in median and mode (which translates into a skewed inclination posterior), they do not necessarily present the most significant differences in the actual orbit shape on the sky plane. Still, the patterns in the posteriors' medians, modes and credible intervals described here hold true and should be kept in mind when estimating the inclination of the orbital plane and its uncertainty.

Another fact that comes to one's attention is that for perfectly edge-on orbits, all three metrics of bias indicate extremely high precision independently of eccentricity and orbital phase. The reasoning behind this is that it is relatively easy to detect a planet moving in an edge-on orbit since it moves in a perfectly straight line from our point of view. However, in order to get such narrow credible intervals as the ones shown in Figure \ref{fig:incbias} we would need a true inclination of exactly $90^\circ$, since very edge-on inclinations like $i_{truth} = 72^\circ$ still exhibit credible intervals similar to other inclinations. Furthermore, we need to remember that even if we had an orbit with exactly $i_{truth} = 90^\circ$, these inclinations are also the cases where the most eccentricity bias in medians and modes, and the widest credible intervals are observed, so the price of perfectly constraining $i$ is an extremely unconstrained posterior in $e$.

The key conclusions regarding the inclination posteriors across orbital phases are:
\begin{itemize}
    \item The performances of the median and the mode at estimating inclination are similar to each other for all eccentricities, inclinations and orbital phases, 
    \item Perfectly face-on inclinations have significant biases in the median and the mode, independently of eccentricity,
    \item The width of the $68\%$ credible interval does not correlate with the biases in the median (even near apastron, the distribution in the Bottom Right panel of Figure \ref{fig:incbias} looks uniform), which hints that inclination appears well-constrained even when it is biased,
    \item As eccentricity increases, less face-on inclinations are affected by significant biases in the median and the mode,
\end{itemize}

\section{The inverse analysis}
\label{sec:inverse}

In all the previous sections we analyzed the biases in the posteriors for known orbital parameters. However, in real life the reason why we need the posteriors is to estimate unknown orbital parameters in the first place. Therefore it is useful to look solely at the \emph{measured} eccentricities and \emph{measured} inclinations and determine which regions of this parameter space are the most reliable or unreliable. In other words, if we get some posteriors for $e$ and $i$, we want to know how certain can we be about the reliability of these estimates.

\begin{figure*}
    \centerline{\includegraphics[width=1.1\textwidth]{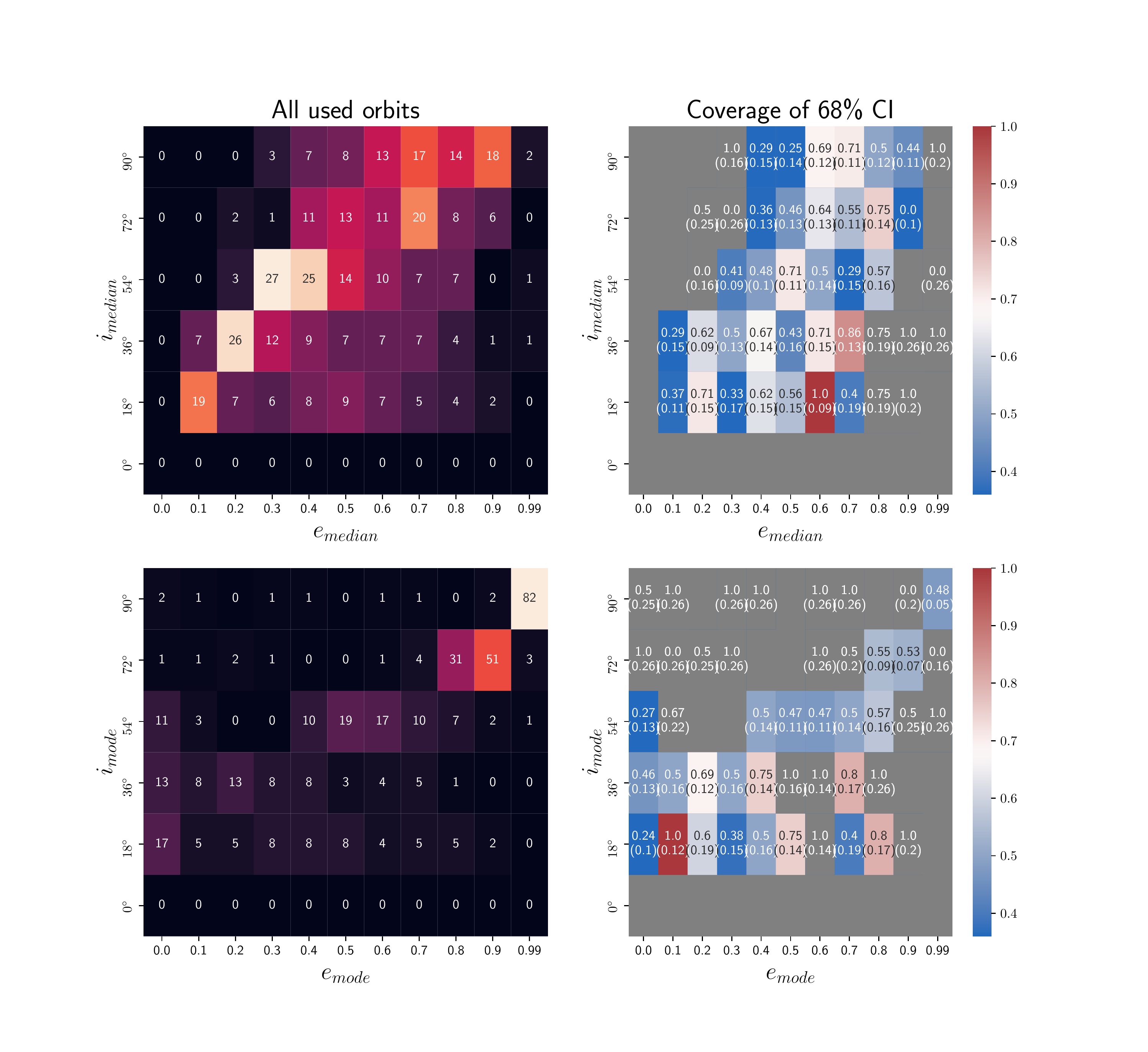}}
    \caption{Measured eccentricities and inclinations. In the left panels, the number in each cell represents how many orbits yielded posteriors with the measured eccentricity given by the $x$-axis ($\pm 0.05$) and measured inclination given by the $y$-axis ($\pm 9^{\circ}$). Top left: All the orbits tested in the 3-D analysis were binned into the 2D histogram according to the median values of $e$ and $i$. Bottom left: Same as Top left but using the 30-bin mode values of $e$ and $i$. Top Right: Effective coverage of the $68\%$ credible interval for $e_{truth}$. The first number in each cell represents the proportion of the orbits in that cell of the Top left panel with $e_{truth}$ inside its $68\%$ credible interval, and the second number (between parenthesis) is the associated uncertainty in the effective coverage, calculated via the beta distribution. The color scheme is centered at $0.68$, so blue cells have effective coverage below $68\%$, and red cells above $68\%$. Bottom right: Same as the Top right, but for the orbits in the Bottom left panel. We see that there is an obvious over-representation of orbits for both the medians and the modes towards the middle of the grids, and a radical under-representation towards the edges. From the panels on the right, we see that $e_{truth}$ tends to be under-covered by the $68\%$ credible interval.}
    \label{fig:hist2d}
\end{figure*}

For this, we decided to bin the studied orbits according to the estimated values of $e$ and $i$. This helps us to determine which regions of parameter space were over-represented and under-represented by the posteriors, understanding an  over-represented cell as one having more orbits than expected ideally, and an under-represented cell as one having fewer orbits than expected ideally. For the binning process, we used all the tested orbits in our three-dimensional analysis. In one case the binning was done according to their median eccentricity and median inclination, and in another to their mode eccentricity and mode inclination. The width of each eccentricity bin is $0.1$ and it is symmetric, so, for example, all the orbits whose measured eccentricity land in the interval $0.45 - 0.55$ will be binned into the $e = 0.5$ bin. Similarly, the height of each cell is $18^{\circ}$ and symmetric, so, for example, any orbit whose measured inclination lands in the interval $9^{\circ} - 27^{\circ}$ will be binned into the $i = 18^{\circ}$ bin. The results are shown in the left-hand side panels of Figure \ref{fig:hist2d}. We remind the reader that we tested $6$ different values of $\tau$, that is, $6$ orbital phases for each pair $(e_{truth}, i_{truth})$. This means that the ideal distribution of the two left-hand side panels would be an uniform 2-D histogram where each cell would have exactly $6$ orbits.

With regards to the credible intervals, we consider a $68\%$ credible interval reliable if the true value lies inside the credible interval $68\%$ of the time after repeating the experiment multiple times. To assess the reliability of the credible intervals, we took the orbits of each individual cell from the histograms in the left panels of Figure \ref{fig:hist2d} and calculated the proportion of those orbits with $e_{truth}$ inside that credible interval. We will refer to this proportion as the effective coverage of the credible interval, since it provides a measure of the frequency with which the true eccentricity is covered by the $68\%$ credible interval. This frequency should ideally be $68\%$, while the credible interval is said to over-cover the true value if the frequency is higher than $68\%$, and under-cover the true value if it is lower than $68\%$. The uncertainty in the effective coverage was calculated using eq. (3) in \citealt{Cameron_2011} (the reported values in Figure \ref{fig:hist2d} are half the width of the $68\%$ credible interval computed with this approach). We chose this method instead of the usual Gaussian approximation used to calculate credible intervals in the binomial distribution because the Gaussian approximation performs poorly at success rates near $0$ or $1$ and when the sample sizes are small, as happens in most of our cells. Only the cells with 5 or more orbits where considered for the analysis of the effective coverage, since cells with 4 or less orbits have too high uncertainties. The results are shown in the right-hand side panels of Figure \ref{fig:hist2d}.

In the case of the medians, we found that there is an evident over-representation of orbits for moderate eccentricities and near-edge-on inclinations. The latter is easy to understand by comparing these results with the ones in Figure \ref{fig:incbias}, where there is a clear preference for near-edge-on inclinations. It is also interesting to note that both circular and extremely eccentric orbits seem to be easily discarded, which is to be expected from the Lucy-Sweeney bias, as we have discussed before. Face-on orbits are also completely under-represented, which is to be expected since Figure \ref{fig:incbias} shows very high bias for both the median and the mode at $i = 0^\circ$ with all tested eccentricities.

With regards to the mode as an estimator, we see a different kind of bias. The first thing that comes to notice is the obvious over-representation for orbits with high eccentricity modes and edge-on inclination modes. The phenomenon of high eccentricities can be more easily visualized in Figure \ref{fig:stacked_hist}. We see that for near-edge-on inclinations the eccentricity posterior tends to form a peak at very high values, and afterwards in Section \ref{sec:analysis2d} we found that this happens independently of the true $e$. From Section \ref{sec:inclination} we can also see that high eccentricities heavily bias inclination towards edge-on values. All together, high eccentricity, edge-on orbits are very over-represented, that is why the upper corners of the mode grid are over-populated.

The absence of face-on orbits in the modes grid is explained by the same reasons as in the medians grid, since, as we discussed in Section \ref{sec:inclination}, the inclination posteriors exhibited medians and modes very close to each other. However, it is worth pointing out that by using the mode as an estimator we can actually measure circular orbits ($e = 0$) with non-near-edge-on inclinations.

In the two right-hand side panels we note a tendency of the effective coverage to be below $68\%$ (that is, the cells on average are generally blue); this suggests that the true value of eccentricity can lie outside the $68\%$ credible interval with more frequency than the expected $32\%$. We shall investigate this further in Section \ref{sec:casestudy}. We do not find, however, a clear dependence of the coverage of the $68\%$ credible interval for eccentricity on the measured eccentricity or the measured inclination.

In practice, Figure \ref{fig:hist2d} gives us an idea of how confident (or dubious) should we be of our eccentricity and inclination posteriors depending on the region our estimates fall in. For example, if we run an orbit-fit and get that our desired orbit has $e_{median} \sim 0.40$ and $i_{median} \sim 54^{\circ}$, we should be wary that many different orbital configurations could produce the same results. On the other hand, if we get, say, $e_{median} \sim 0.20$ and $i_{median} \sim 18^{\circ}$, we can be reasonably confident of our credible interval for eccentricity, because the effective coverage of that cell is well above $68\%$. These grids also tell us that it would be almost impossible to identify orbits with $i_{truth} = 0^{\circ}$ with short orbital arcs. We will discuss this further in Section \ref{sec:casestudy} by looking at more specific cases.
 
The results from this Section also tell us about the practicality of Sections \ref{sec:results} and \ref{sec:inclination}. Even though there are regions of the $e_{truth}$, $i_{truth}$ parameter space where the bias is minimal, these cannot be as easily identified when confronted with real posteriors.

The key conclusions from this section are the following:
\begin{itemize}
    \item The regions of minimal bias described in Section \ref{sec:results} cannot be easily identified when dealing with real posteriors,
    \item The Lower Left panel of Figure \ref{fig:hist2d} suggests that when posteriors show both high $e_{mode}$ and near-edge-on $i_{mode}$, the underlying values of $e_{truth}$ and $i_{truth}$ are essentially unconstrained,
    \item All together, the $68\%$ credible interval of the eccentricity posterior tends to under-cover $e_{truth}$ (that is, the $68\%$ credible interval contains the true value of $e$ less than $68\%$ of the time).
\end{itemize}

\section{Case Studies}
\label{sec:casestudy}
In this section we show some examples of how we might use our results. We present an overview of the inclination-eccentricity degeneracy through a concrete example analyzing a particular pair of estimates $(e, i)$. Additionally, we compare the performance of the median and the mode for different estimated eccentricities when the estimated orbital plane is near-face-on and near-edge-on.

\subsection{Inclination and eccentricity degeneracy}
\label{sec:inceccdeg}

One of the key themes from the three previous sections is the clear degeneracy between eccentricity and inclination. As we have discussed, it is very difficult to differentiate a highly eccentric orbit from a very inclined orbit. This can cause us to mistake circular inclined orbits for eccentric, face-on orbits. We will investigate this degeneracy more concretely with examples.

\begin{figure}
    \centerline{\includegraphics[width=0.5\textwidth]{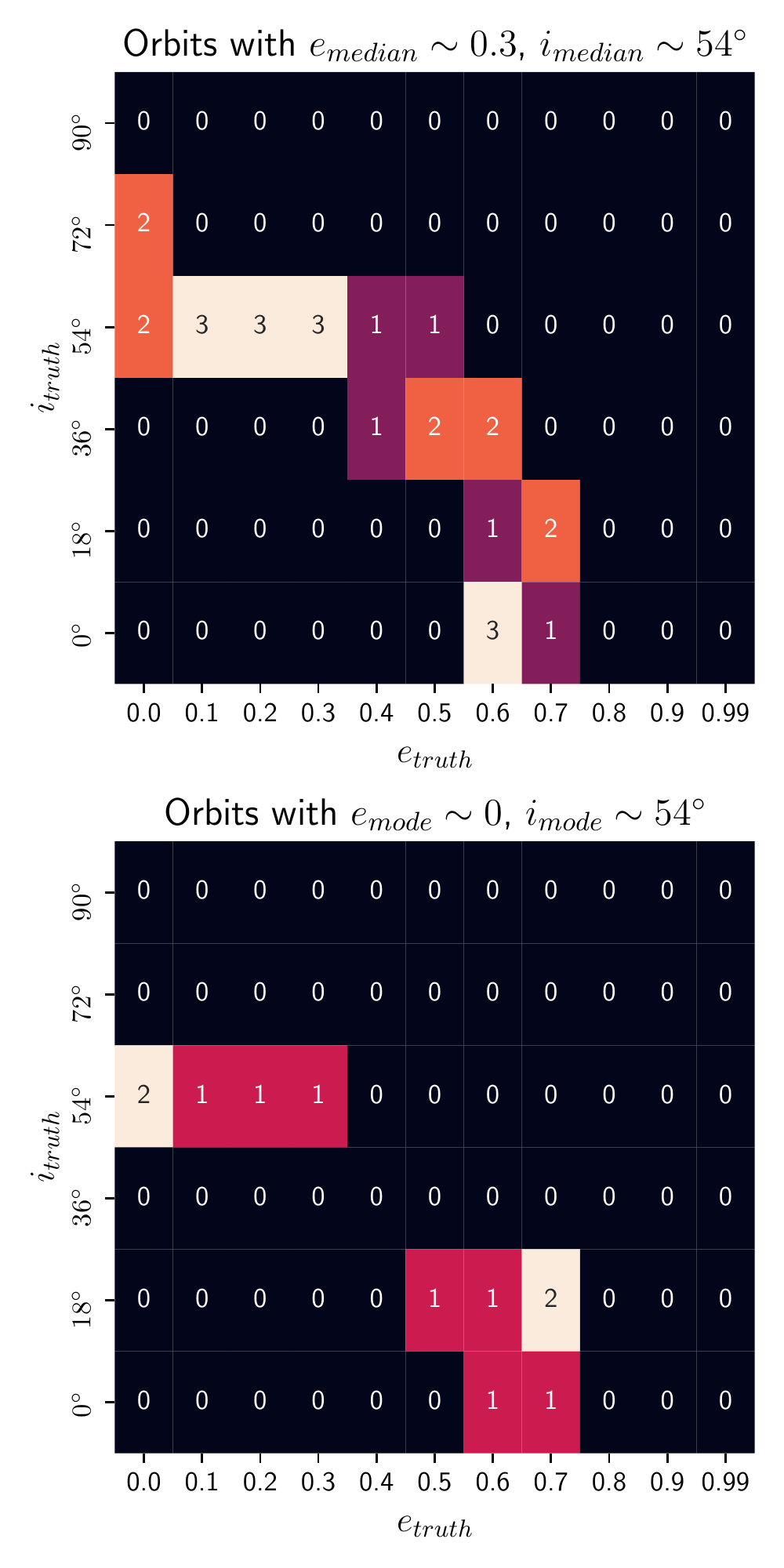}}
    \caption{True eccentricities and true inclinations of orbits whose posteriors match the median estimates of $e$ and $i$ (Top) and the mode estimates of $e$ and $i$ (Bottom) our tested orbit with $e_{truth} = 0$, $i_{truth} = 54^{\circ}$ and $\tau = 0$. The degeneracy of eccentricity and inclination is clearly visible in both cases. Edge-on circular orbits can be easily confused with face-on eccentric orbits.}
    \label{fig:eccincdegen}
\end{figure}

Let's say that we have posteriors that yield values of $e_{median}$ and $i_{median}$ that land on one of the most over represented cells of the Top Left panel of Figure \ref{fig:hist2d} $(e,i) = (0.30, 54^{\circ})$. In that case we can say that our posteriors are most likely very unreliable because there are many orbits that land in that cell, and from the Top Right of Figure \ref{fig:hist2d} panel we see that around $3/5$ of the orbits in that cell have the true $e$ outside their $68\%$ credible interval. To illustrate to what degree can these estimates be a product of bias, in the Top panel of Figure \ref{fig:eccincdegen} we present the true eccentricities and true inclinations of the orbits from our tests that landed in that cell. We see the clear pattern of the $i-e$ degeneracy: we could be dealing either with an eccentric and face-on orbit, with an almost circular and very inclined orbit or something in between.
 
We can do a similar example with $i_{mode}$ and $e_{mode}$. To compare, let's take a specific set of posteriors. From our three-dimensional analysis, we take the posteriors resulting from the orbit $e_{truth} = 0$, $i_{truth} = 54^{\circ}$ and $\tau_{truth} = 0$, whose posteriors yield $e_{median} = 0.30$, $i_{median} = 51.70^{\circ}$, that is, this orbit landed in the previously analyzed cell $(e,i) = (0.30, 54^{\circ})$ of the Top Left panel of Figure \ref{fig:hist2d}. Those same posteriors yielded $e_{mode} = 0.01$ and $i_{mode} = 54.58^{\circ}$, so they would land in the cell $(e, i) = (0, 54^{\circ})$ of the Bottom Left panel of Figure \ref{fig:hist2d}. In the Bottom panel of Figure \ref{fig:eccincdegen} we show the true eccentricities and true inclinations of the orbits whose modes land on the cell $(e, i) = (0, 54^{\circ})$. We see that in fact most of the orbits with $e_{truth} = 0$ and $i_{truth} = 54^{\circ}$ do land on the correct cell, but so do many other undesired orbits. It is also very clear that the $e-i$ degeneracy is present when using the mode as an estimator too, though it seems slightly weaker.

\begin{figure*}
\includegraphics[width=\linewidth]{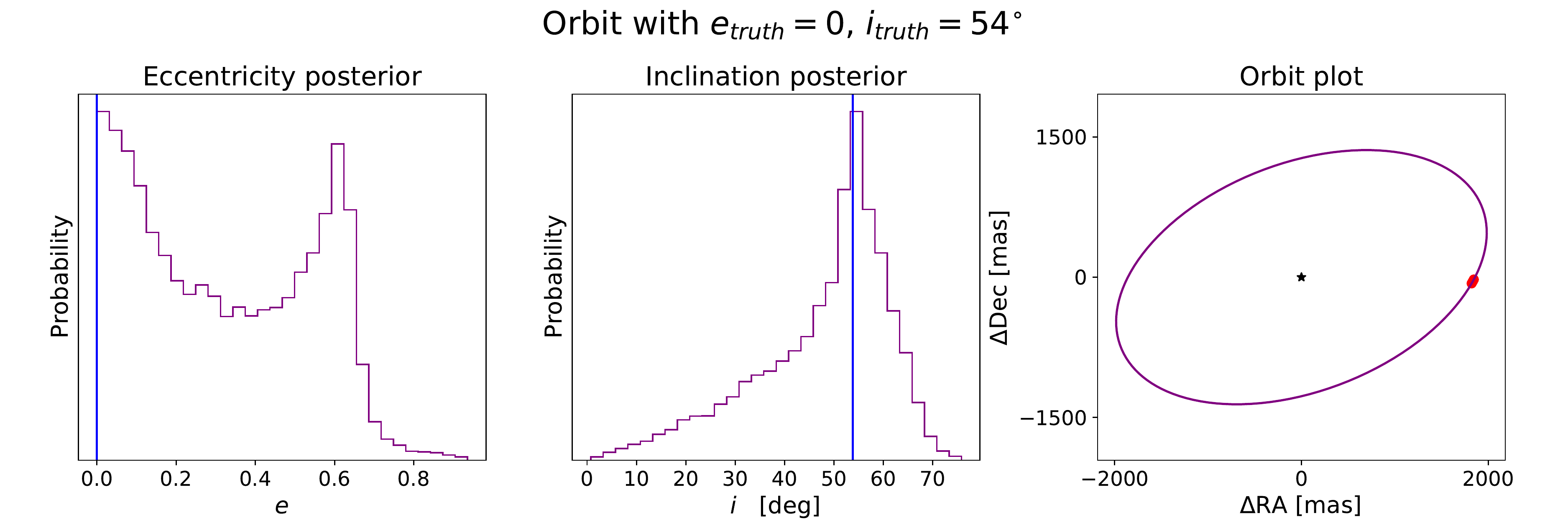}
\includegraphics[width=\linewidth]{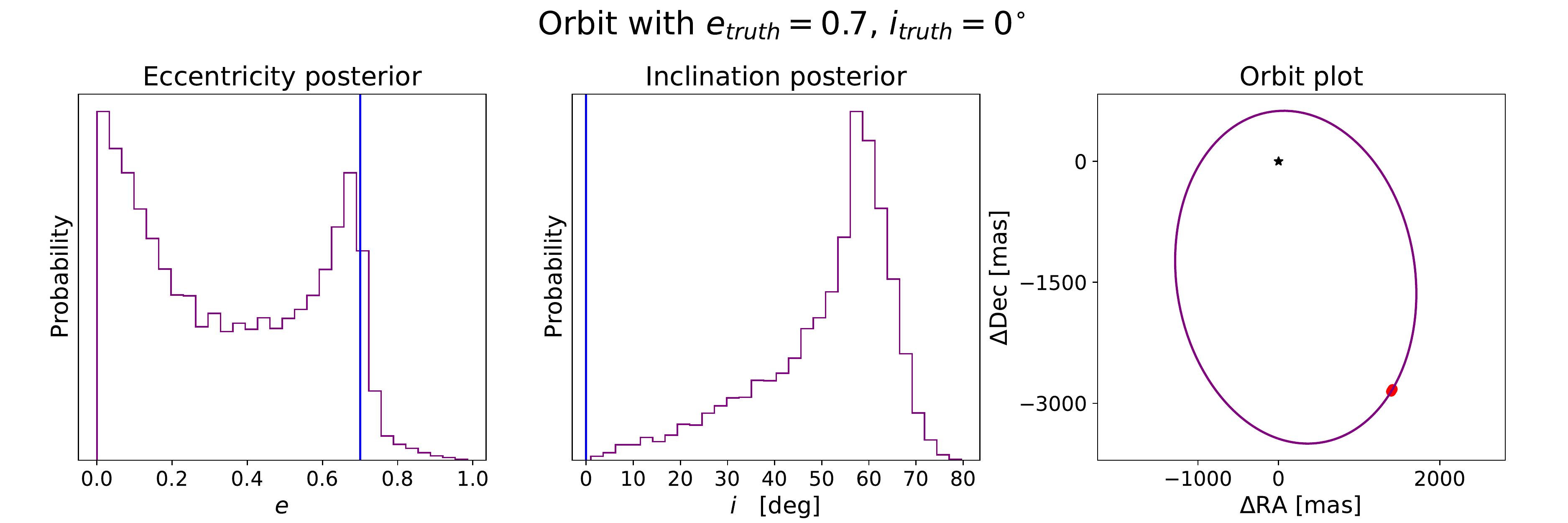}
\caption{Eccentricity and inclination posteriors for two distinct orbits. The top three panels correspond to the first orbit, and the three bottom panels to the second orbit. Their true values are marked with a blue line in the posteriors, and the shape of the true orbit with the position of the observations are plotted in the third panel of each row. We see that not only do the medians coincide in both $e$ and $i$ (as we would expect from Figure \ref{fig:eccincdegen}), but the posteriors' shapes themselves are almost identical.}
\label{fig:twoposts}
\end{figure*}

It can be useful to look at some actual posteriors to better illustrate the extent to which the $e-i$ degeneracy can affect our results. In Figure \ref{fig:twoposts} we see the posteriors resulting from two of our tested orbits: one with $e_{truth} = 0$, $i_{truth} = 54^{\circ}$ and $\tau_{truth} = 0$, and another with $e_{truth} = 0.7$, $i_{truth} = 0^{\circ}$ and $\tau_{truth} = 0$. From the analysis of the previous sections we might expect a behavior like this: a nearly-circular orbit would allow us to see its inclination without much trouble (unless it is exactly face-on, as we said before), but if it is very inclined the eccentricity posterior would be very biased; on the other hand, if our orbit is very eccentric, the inclination would be significantly biased towards more edge-on values. That is, biased posteriors for $e$ and $i$ take place whenever $e_{truth}$ is high or $i_{truth}$ is near-edge-on, and the results from those can be indistinguishable. Not only are the modes and the medians degenerate, but the shapes of both posteriors look almost identical, even though they correspond to radically different orbits. It is interesting to note that both eccentricity posteriors in Figure \ref{fig:twoposts} show peaks at the desired values, but the inclination posteriors only suggest a near-edge-on possibility for inclination. 

\subsection{Different eccentricities at face-on measured inclinations}
\label{sec:diffecc}

In Section \ref{sec:analysis3d} we saw that in the cases of low inclinations the mode of the eccentricity posterior could be a better estimator than the median. However, as we just saw, given the $e-i$ degeneracy some orbits in certain regions of the $e_{truth}$, $i_{truth}$ parameter space can end up in very distant regions in the measured $e$, measured $i$ parameter space. Therefore, it might be a good idea to take a look at these regions of the measured parameter space more closely and compare the performances of the median and the mode.

\begin{figure*}
    \centerline{\includegraphics[width=0.9\textwidth]{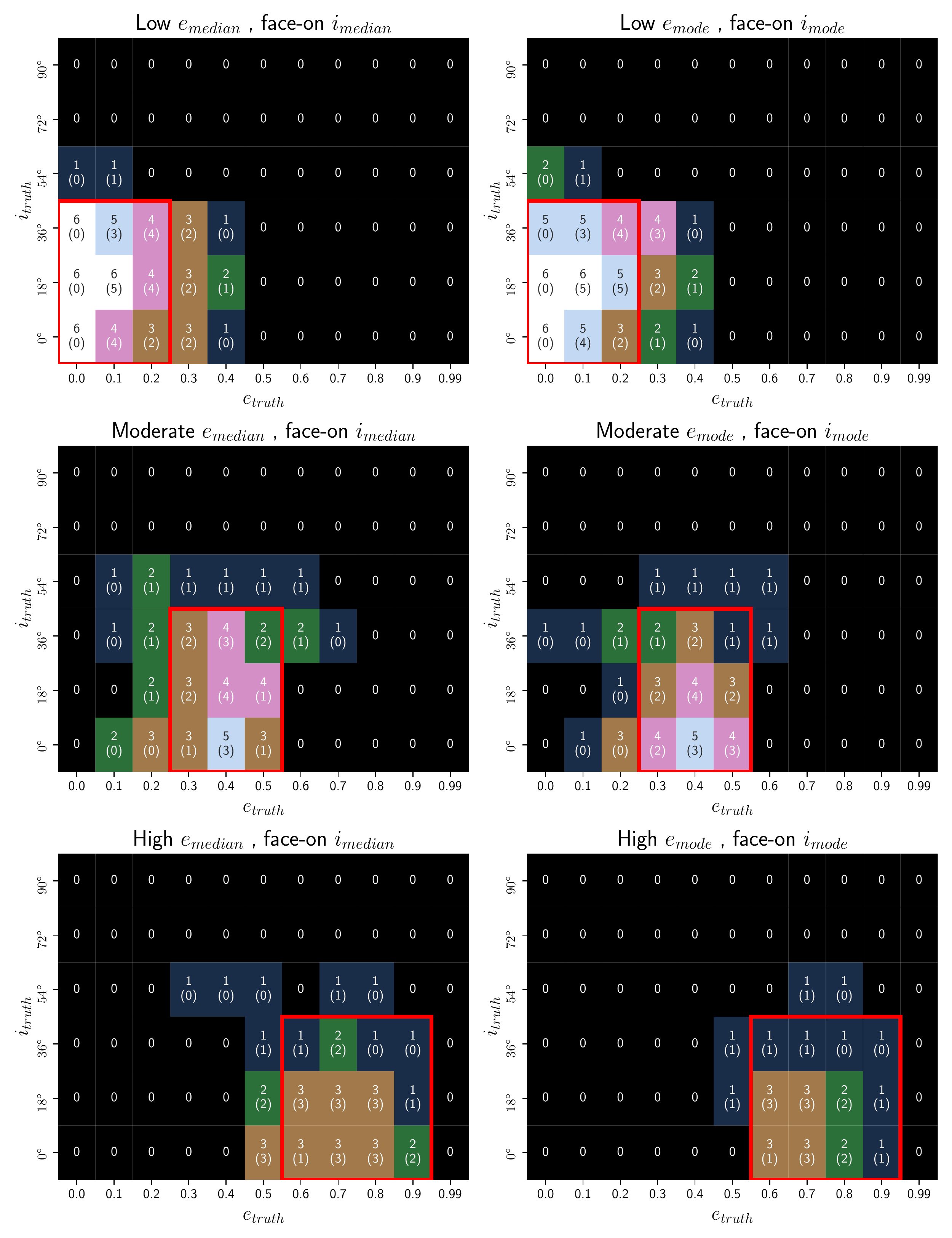}}
    \caption{True orbital parameters of orbits with the specified measured values of $e$ and $i$. For the three left-hand side panels, the red rectangles indicate the cells of the Top Left panel of Figure \ref{fig:hist2d} from which orbits were drawn. For the three right-hand side panels, it was from the Bottom Left panel of Figure \ref{fig:hist2d}. The first number on each cell indicates the number of orbits with those $e_{truth}$, $i_{truth}$ that landed in the red rectangle of the corresponding measured parameter space. The numbers in parenthesis in each cell indicate the number of orbits from that cell with $e_{truth}$ inside the $68\%$ credible interval of its eccentricity posterior. We do not see a big difference in the median and the mode. In general, the true values are close to the estimates, and the $i-e$ degeneracy does not seem to have a big impact when $i$ is measured to be near face-on.}
    \label{fig:manyecclowinc}
\end{figure*}

For this, while keeping the binning as in Figure \ref{fig:hist2d}, we looked at three regions of median and mode parameter spaces: low measured eccentricity (cells with $e$ from $0.0$ to $0.20$), moderate measured eccentricity ($e$ from $0.30$ to $0.50$) and high measured eccentricity ($e$ from $0.60$ to $0.90$), all with near-face-on measured inclinations (cells with $i$ from $0^\circ$ to $36^\circ$). Then we took the true orbital parameters from those orbits and made a histogram in $e_{truth}$, $i_{truth}$ space. The results are shown in Figure \ref{fig:manyecclowinc}, the coloring and first number of the cells representing the number of orbits with those true orbital parameters that landed in the desired region of measured parameter space (marked with a red rectangle for reference) and the number in parenthesis on each cell indicating the number of those orbits that had their $e_{truth}$ value within their respective $68\%$ credible interval centered at the median.

We find that in all cases the distribution of true orbital parameters is similar in the same regions of median parameter space and mode parameter space. No big differences are noticed between median and mode, aside from the fact that for moderate and high measured eccentricities the true orbital parameters seem to be slightly closer to the measured region. However, one important thing to note is that most orbits do land inside the red boxes, which means that in this measured inclination regime we can at least differentiate between low, moderate and high eccentricity orbits.

To assess quantitatively the difference in the performance of the median and the mode in each of these regions of measured parameter space, we computed the biases in the median and the mode as defined previously (i.e. $\beta_{med} = \lvert e_{median} - e_{truth} \rvert$, $\beta_{mode} = \lvert e_{mode} - e_{truth} \rvert$) for all the orbits, and then we calculated the median of each bias for the orbits in each region separately. Additionally, we calculated the widths of the $68\%$ credible interval ($w_{68}$) and calculated the median for the orbits in each region. Finally, we counted the total number of orbits that landed in each region and the number of orbits from each region with $e_{truth}$ within its corresponding $68\%$ credible interval. The results are shown in the face-on measured inclinations of Table \ref{tab:biasesregs}.

\begin{deluxetable*}{llccccc}
\tablecaption{Biases and effective coverage in eccentricity for different regions of measured eccentricity and inclination$^a$ \label{tab:biasesregs}}
\tablehead{\colhead{Measured $e$} & \colhead{Measured $i$} & \colhead{Total orbits} & \colhead{Proportion with $e_{truth}$ within $68\%$ CI}&\colhead{med$\beta_{med}$} & 
\colhead{med$\beta_{mode}$} & \colhead{med$w_{68}$}}
\startdata
Low $e_{median}$ & Face-on $i_{median}$& $59$ & $50.84\%$ & $0.07$& $0.05$ & $0.24$\\
Low $e_{mode}$ & Face-on $i_{mode}$ & $61$ & $50.81\%$ & $0.08$ & $0.05$ & $0.24$ \\ 
Moderate $e_{median}$ & Face-on $i_{median}$& $51$ & $52.94\%$& $0.10$ &$0.07$ & $0.27$\\
Moderate $e_{mode}$& Face-on $i_{mode}$& $43$ & $60.46\%$ & $0.08$ & $0.05$ & $0.25$\\
High $e_{median}$ & Face-on $i_{median}$ & $37$ & $78.37\%$ & $0.05$ & $0.03$ & $0.17$\\
High $e_{mode}$ & Face-on $i_{mode}$& $26$& $80.76\%$ & $0.04$ & $0.03$ & $0.15$\\
Low $e_{median}$ & Edge-on $i_{median}$& $5$ & $20\%$ & $0.22$& $0.05$ & $0.59$\\
Low $e_{mode}$ & Edge-on $i_{mode}$ & $21$ & $42.85\%$ & $0.22$ & $0.18$ & $0.53$ \\ 
Moderate $e_{median}$ & Edge-on $i_{median}$& $109$ & $45.87\%$& $0.25$ &$0.14$ & $0.56$\\
Moderate $e_{mode}$& Edge-on $i_{mode}$& $32$ & $53.12\%$ & $0.15$ & $0.08$ & $0.37$\\
High $e_{median}$ & Edge-on $i_{median}$ & $131$ & $54.19\%$ & $0.14$ & $0.09$ & $0.52$\\
High $e_{mode}$ & Edge-on $i_{mode}$& $127$& $52.75\%$ & $0.18$ & $0.07$ & $0.52$
\enddata
\tablenotetext{a}{Low, moderate and high measured eccentricities correspond to values of $e = 0 - 0.2$, $e = 0.3 - 0.5$, $e = 0.6 - 0.9$, respectively, as binned in Figure \ref{fig:hist2d}. Face-on and edge-on measured inclinations correspond to values of $i = 0^\circ - 36^\circ$ and $i = 54^\circ - 90^\circ$, respectively, also as binned in Figure \ref{fig:hist2d}. The fourth column is the proportion of orbits from the total with values of $e_{truth}$ that land within the $68\%$ credible interval (centered at the median) of their corresponding posterior. The used statistics in the last three columns correspond to bias in the median ($\beta_{med} = \lvert e_{truth} - e_{median} \rvert $), bias in the mode ($\beta_{mode} = \lvert e_{truth} - e_{mode} \rvert $) and width of the $68\%$ credible interval centered at the median $w_{68}$. The reported values correspond to the median of the statistics calculated across the orbits in the specified region.}
\end{deluxetable*}

In general, the credible intervals are relatively narrow, so most posteriors are probably not bi-modal. We see too that the biases in the median at face-on measured inclinations are always very small (most of them lower than $0.1$). We  also notice that the differences in the median biases of the median and the mode are not huge, but the median bias in the mode is consistently lower than the one in the median. We also see that the median widths of the $68\%$ credible intervals are very similar in the same regions of median and mode parameter spaces. With these results, we find evidence that the mode could be a better estimator of eccentricity than the median when measured inclination is near-face-on, but the differences are minimal. The number of orbits in each region is also similar for median and mode parameter spaces; the biggest difference was at high measured eccentricities, where the total number of orbits in median space is $\sim$ $40\%$ larger than the total number of orbits in mode space.

The fourth column of Table \ref{tab:biasesregs} conveys important information about the coverage of the $68\%$ credible interval for eccentricity: it can be seen that the proportion of orbits with $e_{truth}$ within the credible interval is almost always below $0.68$, the two exceptions being at high measured eccentricities. We interpret this as evidence that the $68\%$ credible interval for eccentricity under-cover $e_{truth}$ in most cases, a conclusion that has been pointed out in previous studies with uninformative priors on the Keplerian orbital elements (e.g. \citealt{O_Neil_2019}, \citealt{Lucy_2014}). Additionally, we find no significant differences in the effective coverage when using the median as an estimator compared to using the mode as an estimator.

In summary, we found that for near-face-on measured inclinations the $i-e$ degeneracy does not greatly affect our ability to estimate eccentricity or inclination, since both the median and mode estimates are usually reasonably close to the true orbital elements for most measured eccentricities. We also found that the eccentricity credible intervals are not notably wide in this regime of measured inclinations, which means that most eccentricity posteriors are not bi-modal or particularly skewed. Finally, we found evidence that at near-face-on measured inclinations the mode is a slightly better estimator of eccentricity than the median, but the differences are not very big.

\subsection{Different eccentricities at near-edge-on measured inclinations}
\label{sec:highinc}

\begin{figure*}
    \centering{\includegraphics[width=0.9\textwidth]{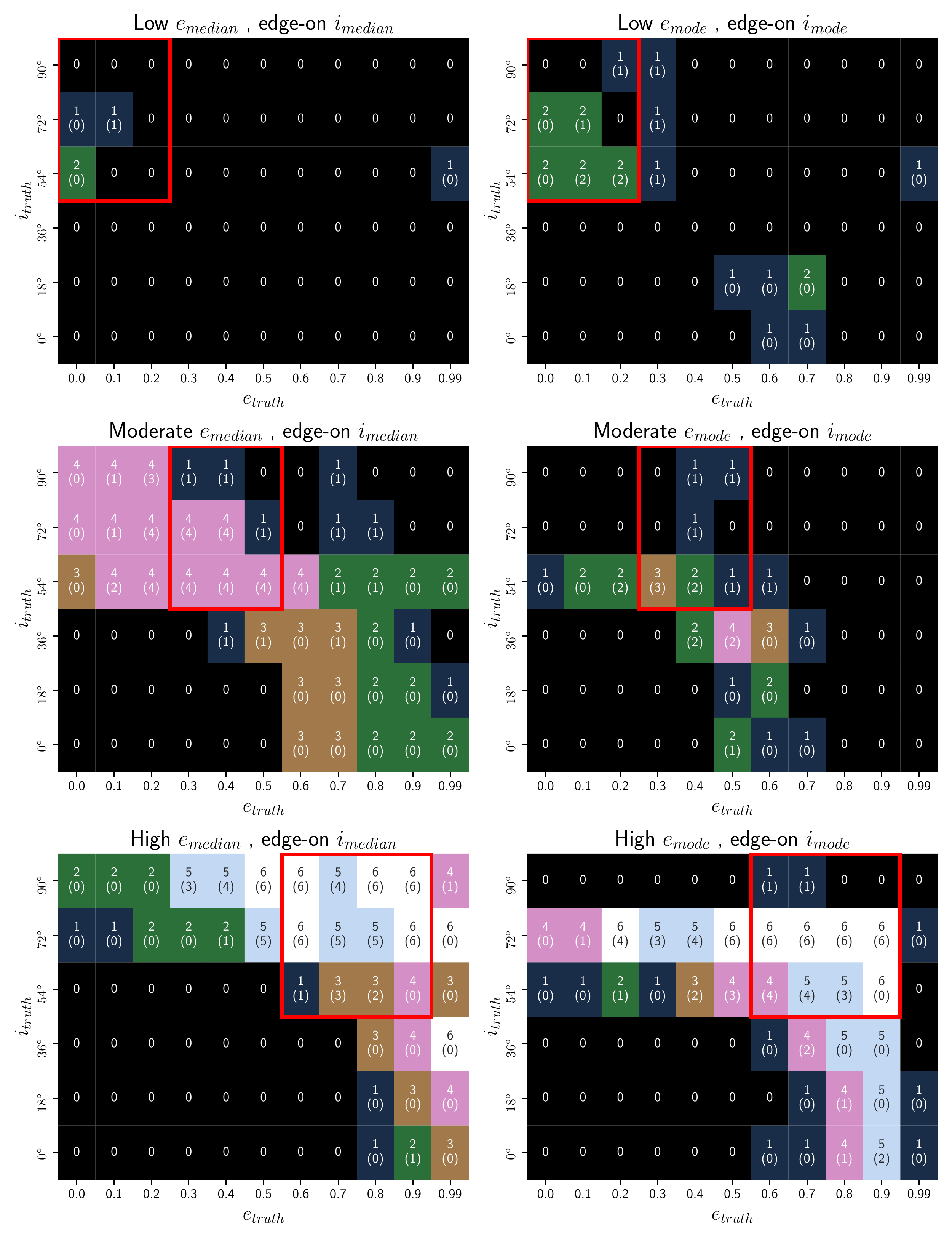}}
    \caption{Same as Figure \ref{fig:manyecclowinc} but for near-edge-on measured inclinations. We see that the degeneracy between eccentricity and inclination becomes much more significant in this regime. By comparing this figure with Figure \ref{fig:manyecclowinc}, we see that as long as our \emph{measured} inclination is near-face-on, our estimates will in general be reasonably close to the true orbital parameters.}
    \label{fig:manyecchighinc}
\end{figure*}

To compare, we perform the same analysis as in the previous subsection but now with near-edge-on measured inclinations ($i$ from $54^\circ$ to $90^\circ$). We summarize the results in Figure \ref{fig:manyecchighinc}, and in the edge-on measured inclinations of Table \ref{tab:biasesregs}.

It is easy to see that the $i-e$ degeneracy has a much stronger effect in this regime than in near-face-on measured inclinations. This tells us to be very wary of our estimates whenever $i_{mode}$ or $i_{median}$ is near-edge-on, independently of the measured eccentricity.

In the edge-on measured inclination entries of Table \ref{tab:biasesregs} we see that for all regions the width of the $68\%$ credible intervals are huge compared to the face-on measured entries. All the biases in the median and in the mode are also larger overall than their face-on inclination counterparts. Finally, we also note that in moderate and high $e_{median}$ the total number of orbits is considerably bigger, and that low values of $e_{median}$ barely have any orbits; this, along with the previous remarks about the wide credible intervals, indicates that the regions with near-edge-on \emph{measured} inclinations have very unconstrained eccentricity.

In the top two panels of Figure \ref{fig:manyecchighinc} we see many more orbits in the mode parameter space than in the median, but looking back at Figure \ref{fig:hist2d} we see that it is to be expected: the chosen region encompasses the cell $(e_{mode}, i_{mode}) = (0, 54^\circ)$, which has distinctively more orbits than its neighbor cells in mode space. 

The biggest difference that we find is in the middle panels of Figure \ref{fig:manyecchighinc}, where we see that orbits with moderate $e_{median}$ span a much broader region of the parameter space of true orbital elements than orbits with moderate $e_{mode}$. We also find that many cells have $100\%$ of their orbits with $e_{truth}$ within the $68\%$ credible interval (i.e. there are many cells with the same number inside and outside the parenthesis), but this may not be useful since the credible intervals are very wide in the first place.

With regards to the region of high measured eccentricities, we find that for both the median and the mode there are a great number of orbits. One important fact is the complete absence of orbits with $i_{truth}=90^\circ$ in mode space: most of them land in the top right corner of the grid $(e_{mode}, i_{mode}) = (0.99, 90^\circ)$, as we can see from the Bottom Left panel of Figure \ref{fig:hist2d} and the shape of the $i=90^\circ$ eccentricity posterior of Figure \ref{fig:stacked_hist}. This tells us that $e_{truth}$ is completely unconstrained for perfectly edge-on orbits.

As in the previous section, we find that, this time without exception, the proportion of orbits with $e_{truth}$ inside the $68\%$ credible interval is consistently lower than $0.68$. So despite the very wide credible intervals that the eccentricity posterior tends to have in this regime, these still cover the true value less that $68\%$ of the time. Also, like in the previous case, we do not find a significant difference in the proportion of orbits between median and mode spaces.

In summary, when measured inclination is near-edge-on, the $i-e$ degeneracy greatly harms our ability to estimate $e_{truth}$ and $i_{truth}$ independently of eccentricity, and eccentricity is generally unconstrained, with no significant differences in the performance of the median or the mode at estimating eccentricity.

The main takeaways from this section are:
\begin{itemize}
    \item There is a degeneracy that makes posteriors from highly-eccentric, near-face-on orbits very difficult to distinguish from near-circular, near-edge-on orbits,
    \item Inclination posteriors can be highly biased (with the median and mode far apart from the underlying $i_{truth}$) but still appear well-constrained (not bi-modal and with relatively narrow $68\%$ credible intervals),
    \item The $68\%$ credible intervals of eccentricity of orbits tested with either $i_{med}$ or $i_{mode}$ near-edge-on showed an effective coverage consistently below $0.68$, while orbits with either $i_{med}$ or $i_{mode}$ near-face-on showed an effective coverage mostly below $0.68$.
\end{itemize}

\section{Conclusions and recommendations}
\label{sec:recommendations}
In this study we presented an analysis on biases in the posteriors of orbital elements for directly-imaged exoplanets when only a very small fraction of the orbital period is observed. We performed orbit-fits with mock data from a range of orbits and assessed patterns from the results. We focused on eccentricity and inclination since we found that those two elements were the ones with the greatest impact in the eccentricity posterior.

Considering our results as a whole, we present the following main takeaways. Once again, we remind the reader that these conclusions hold in cases where the observed orbital arc is very small ($\sim 0.5 \%$), and not all of them might hold otherwise.
\begin{itemize}
    \item Near-edge-on true inclinations produce bi-modal eccentricity posteriors.
    \item For high true values of eccentricity, inclination posteriors tend to be heavily biased towards more edge-on values.
    \item When the observed orbital arc is too small (so small that no changes in the planet speed are detected) one more data point does not significantly improve the eccentricity posteriors.
    \item The accuracy of the mode to estimate eccentricity eccentricity depends almost purely on the inclination: it is a good estimator for any value of eccentricity if the measured inclination is near-face-on. There is also evidence that in this inclination regime the mode gives a better estimate than the median.
    \item There is a clear degeneracy between eccentricity and inclination: it is hard to distinguish a very eccentric and face-on orbit from a near-circular and very inclined orbit.
    \item When the inclination posterior suggests a near-face-on inclination, the effect of the degeneracy between eccentricity and inclination is very small, and thus the eccentricity estimates tend to be close to the true value (except for perfectly circular orbits, due to the well-known Lucy-Sweeney bias). As a consequence, we can distinguish between low, moderate and high eccentricity orbits.
    \item When the inclination posterior suggests a near-edge-on inclination, the effect of the $i-e$ degeneracy is much more significant, and eccentricity will be essentially unconstrained.
    \item Figure \ref{fig:hist2d} and Table \ref{tab:biasesregs} suggest that the $68\%$ credible intervals for eccentricity almost always under-cover the true value (i.e. $e_{truth}$ lands within the $68\%$ credible intervals less than $68\%$ of the time), so we recommend caution when interpreting orbital constraints from the $68\%$ credible interval for eccentricity.
\end{itemize}

The orbit-fits performed in this study are comparable to long-period companions with less than $\sim 1 \%$ period coverage, like HD 23514 B, 2M1559+4403 B and 1RXS2351+3127 B, according to the best-fit orbital periods reported by \citet{Bowler_2020}. Thus, future work could be aimed at studying biases in orbit-fitting when a larger portion of the orbital arc is observed, like 51 Eridani b \citep{DeRosa_2019}, to determine which of the biases found in this study remain and which ones are resolved, as well as how much orbital coverage is needed for the $e-i$ degeneracy to be mitigated given a certain astrometric precision (in this study, we used uncertainties of $1$ $\mathrm{mas}$ in both $\Delta \mathrm{RA}$ and $\Delta \mathrm{Dec}$). Another possible line of work would be to further investigate the effects of choosing different priors when performing orbit-fits, or constructing minimally biasing priors (e.g. \citealt{O_Neil_2019}). Finally, if both RV data and astrometry were available, like the cases of $\beta$ Pictoris b \citep{Snellen_2014}, HR 8799 b and c \citep{Ruffio_2019} and GQ Lupi b \citep{Schwarz_2016}, it could be investigated how much RV data would be necessary to resolve the biases presented in this work.

\acknowledgments
This work was supported by the Heising-Simons Foundation through grant 2019-1698. We thank Kelly O'Neil, Rob De Rosa, and Isabel Angelo for their insights. We also thank the anonymous referee, whose comments greatly improved the quality of this work. 

\bibliographystyle{aasjournal}
\bibliography{ref}

\end{document}